\documentclass[aps,prl,twocolumn, showpacs, nobibnotes, superscriptaddress, bibliography]{revtex4-2}
\usepackage{amsbsy,amssymb,amsmath,bm}
\usepackage{graphicx}
\usepackage{braket}
\usepackage{float}
\usepackage{textcomp}
\usepackage{color}
\usepackage[colorlinks=true,linkcolor=red]{hyperref}
\usepackage[sort&compress]{natbib}
\usepackage[normalem]{ulem}
\usepackage[shortlabels]{enumitem}
\usepackage{appendix}
\usepackage[table]{xcolor}
\usepackage{xcolor,colortbl}
\usepackage{soul}
\usepackage{amsmath}
\usepackage{mhchem}

\usepackage[
  nameinlink,
  capitalize]{cleveref}
\crefname{equation}{Eq.}{Eqs.}
\Crefname{equation}{Equation}{Equations}
\crefname{figure}{Fig.}{Figs.}
\Crefname{figure}{Figure}{Figures}
\usepackage{siunitx}

\input{epsf}

\newcommand{\tokyotech}{Department of Electrical and Electronic Engineering, School of Engineering, Institute of Science Tokyo, 2-12-1 Ookayama, Meguro, Tokyo 152-8552, Japan}
\newcommand{\nims}{National Institute for Materials Science, 1-1 Namiki, Tsukuba, Ibaraki 305-0044, Japan}

\begin{document}
\title {Observation of individual vortex penetration in a coplanar superconducting resonator}

\author{Kirill Shulga}
\email{kirill\_shulga@protonmail.ch}
\affiliation{International Center for Elementary Particle Physics, The University of Tokyo, 7-3-1 Hongo, Bunkyo-ku, Tokyo 113-0033, Japan}
\author{Shunsuke Nishimura} 
\affiliation{Department of Physics, The University of Tokyo, Bunkyo-ku, Tokyo 113-0033, Japan}
\author{Pavel A. Volkov}
\affiliation{Department of Physics, University of Connecticut, Storrs, Connecticut 06269, USA}
\author{Miu Hirano}
\affiliation{Department of Physics, University of Arizona, Tucson, AZ, USA, 85721}
\author{Toshiaki Inada}
\affiliation{International Center for Elementary Particle Physics, The University of Tokyo, 7-3-1 Hongo, Bunkyo-ku, Tokyo 113-0033, Japan}
\author{Ryota Hasegawa}
\affiliation{Department of Physics, The University of Tokyo, Bunkyo-ku, Tokyo 113-0033, Japan}

\author{Takeyuki Tsuji}
\affiliation{\tokyotech}
\affiliation{%
 \nims
}%

\author{Takayuki Iwasaki}
\affiliation{\tokyotech}

\author{Mutsuko Hatano}
\affiliation{\tokyotech}
\author{Kento Sasaki}
\affiliation{Department of Physics, The University of Tokyo, Bunkyo-ku, Tokyo 113-0033, Japan}
\author{Kensuke Kobayashi}
\affiliation{Department of Physics, The University of Tokyo, Bunkyo-ku, Tokyo 113-0033, Japan}

\date {\today}

\begin{abstract}
We demonstrate the microwave signatures of individual Abrikosov-vortex penetration events in superconducting microwave resonators. $\lambda/4$ resonators with a narrowed region near the grounded end acting as a vortex trap were fabricated and studied using microwave transmission spectroscopy at millikelvin temperatures. Sharp stepwise drops in resonance frequency are detected as a function of increasing external magnetic field, attributed to the entry of individual Abrikosov vortices in the narrow region. This interpretation is confirmed by NV center magnetometry revealing discrete vortex entry events on increasing field and the absence of such drops in a device with a wider narrowed region.
Our results establish a method to investigate and manipulate the states of Abrikosov vortices with microwaves access and readout vortex configurations using magnetic-field sweeps and microwave drive.
\end{abstract}

\maketitle
\textit{Introduction.—}
Vortices are the elementary topological defects of superconductors, which have been in recent focus for both the fundamental insights their properties can provide and their applications.
Their behavior encodes detailed information about the pairing symmetry~\cite{matsumoto1999chiral,eschrig2009charge,liu2024}, pinning landscape, and vortex-core excitations~\cite{drew_1992,choi1994,eschrig1999electromagnetic,PhysRevB.60.9295}, the latter having been shown to realize Majorana modes in appropriate settings~ \cite{fukane,beenakker2013search}. 
In addition, field-controlled access to vortex configurations has been proposed as a route toward nonvolatile memory elements compatible with superconducting quantum electronics~\cite{kalashnikov2024demonstration, Foltyn2024}. These developments motivate experimental architectures capable of resolving the electrodynamic response of individual vortices with high sensitivity. Discrete changes of the superconducting state upon vortex entry and exit are a central theme of mesoscopic superconductivity, as established by magnetometry of individual fluxoid states and by Ginzburg--Landau studies of mesoscopic vortex states~\cite{Geim2000Fluxoid,Deo1997MesoscopicDisks}. Our work transfers this few-vortex physics to a microwave-resonator geometry, where an engineered current antinode converts individual vortex entry into a dispersive microwave signal.

Superconducting planar resonators provide such a platform. Their high quality factors and narrow linewidths make them exceptionally sensitive probes of electromagnetic and quantum phenomena at the microscale~\cite{Zmuidzinas2012, Blais2021, Gu2017, Krantz2019, Day2003, Zmuidzinas2004, Göppl2008}. Magnetic flux penetration resulting in a  finite density of Abrikosov vortices (AVs) alters both the reactive and dissipative response of coplanar resonators \cite{Song2009, Bothner2011, Stan2004, Clem2011} and may lead to flux jumps - instabilities of the bulk superconducting state \cite{mints1981}. However, real-time detection of individual vortex events in coplanar waveguide (CPW) resonators remains rare. 
Single-vortex readout has mainly relied on magnetic means, such as scanning probes-scanning SQUID~\cite{Kirtley1999}, scanning Hall~\cite{Chang1992}, MFM manipulation~\cite{Auslaender2009}, including its use in long Josephson junctions~\cite{dremov2019local,grebenchuk2020observation,hovhannisyan2021lateral,stolyarov2022revealing}, Lorentz-TEM~\cite{Harada1992}, Magneto-Optic microscope~\cite{Veshchunov2016, nulens2023catastrophic}, and nanothermometry \cite{Foltyn2024}, some of which offer spatial resolution but are slow and not easily multiplexed. Using microwaves to detect and read out vortices is a highly promising alternative, having potential advantages for the probing and readout of Majorana states~\cite{vayrynen_2024} (with respect to the scanning probe approach~\cite{stm_rev,stm_maj,musr_rev}), and enabling microwave-assisted access to vortex configurations~\cite{vortcool}. However, detecting and reading out individual vortices with microwaves is intrinsically challenging because the impedance change from a single flux quantum is much smaller than intrinsic surface resistance and background inductive nonlinearities, making enhanced current-density regions essential for resolving discrete vortex-induced shifts. Previously, using variable-width CPW geometries, individual trapped-vortex signatures were observed only after field-cooling~\cite{Nsanzineza2014}, with the response  of individual vortex entry dominated by the dissipative part (quality factor reduction).

In this Letter, we show that engineering a narrow lithographic constriction at the grounded end of a minimal $\lambda/4$ CPW resonator enables field-controlled access to metastable vortex configurations, such that individual penetration events induce sharp, quantifiable steps in the frequency $f_0$ and the quality factor $Q$ of the resonator during a continuous field ramp. This provides a microwave-based form of magnetic single-vortex detection using standard transmission spectroscopy. We directly visualize the discrete vortex entry events  
with a quantum diamond microscope (QDM) employing a perfectly aligned ensemble of nitrogen-vacancy (NV) centers~\cite{SN2023, Tsuji2022}, demonstrating how geometric current concentration in the constriction guides and enhances vortex entry in engineered superconducting structures.

\begin{figure}[h!]
\noindent\centering{
\includegraphics[width=0.47\textwidth]{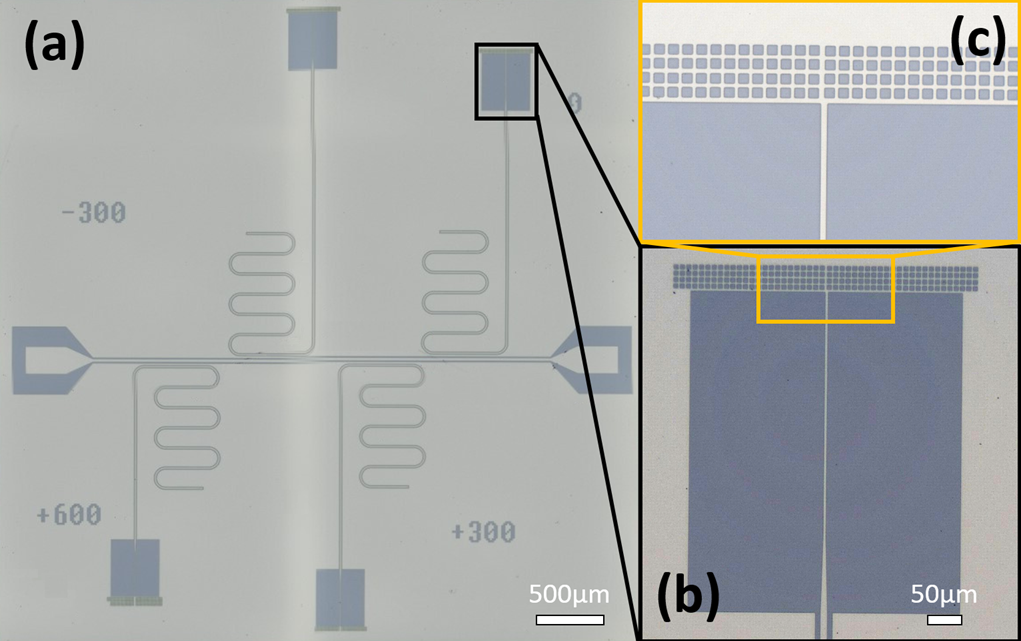}
}
\caption{(a) Optical image of the chip showing the common feedline and four multiplexed $\lambda/4$ resonators. (b) Zoom in on the grounded end of one resonator, showing the transition (“taper region”) between the wide CPW and the thin microstrip constriction (“neck”) in the center conductor. The boxed Nb-etched-out window surrounds the neck and has been shown to produce flux-focusing 
in this region~\citep{Brandt2005}. (c) Further magnification of the 1~$~\si{\micro m}$-wide neck used to pin individual vortices.
The narrow neck at the grounded end is designed to enhance current density and increase sensitivity to vortex entry.}
\label{fig1}
\end{figure}

\noindent\textit{Materials and Methods.—}
The chip used in this experiment features four quarter-wavelength ($\lambda/4$) coplanar waveguide (CPW) resonators capacitively coupled to a standard transmission line, see Fig.~\ref{fig1}. Each resonator was terminated at the grounded end by a narrowed section (or ``neck’’), designed to enhance current density and increase sensitivity to local perturbations such as vortex entry. The constriction had a fixed length of 325~$~\si{\micro m}$ and widths varying from 1 to 10~$~\si{\micro m}$ across devices. The total resonator lengths ranged from 6.4 to 7.2 mm, corresponding to designed variations in the fundamental frequency (4.18–4.7 GHz) for multiplexed readout.

The resonators were fabricated on high-resistivity silicon substrates using 200~nm-thick Nb films deposited via DC magnetron sputtering and patterned by optical lithography and reactive ion etching. The measured loaded quality factor $Q_L$ exceeded $10^5$ without vortices.
We applied a slowly swept out-of-plane magnetic field to probe vortex dynamics using a superconducting coil mounted around the sample.
The magnetic field was swept at a rate below 1 $~\si{\micro T/s}$, ensuring that vortices nucleate and pin quasi-statically without dynamic depinning events.

The microwave measurements were performed at a base temperature of 20~mK in a dilution refrigerator, with magnetic shielding provided by a two-layer $\mu$-metal can and an inner superconducting shield. This ensured sub-$~\si{\micro T}$ residual field levels and reproducible vortex nucleation conditions. After thermal cycling, the exact values of the magnetic field corresponding to vortex entry varied, but the step heights in frequency remained consistent between runs.

\begin{figure}[h!]
\noindent\centering{
\includegraphics[width=84mm]{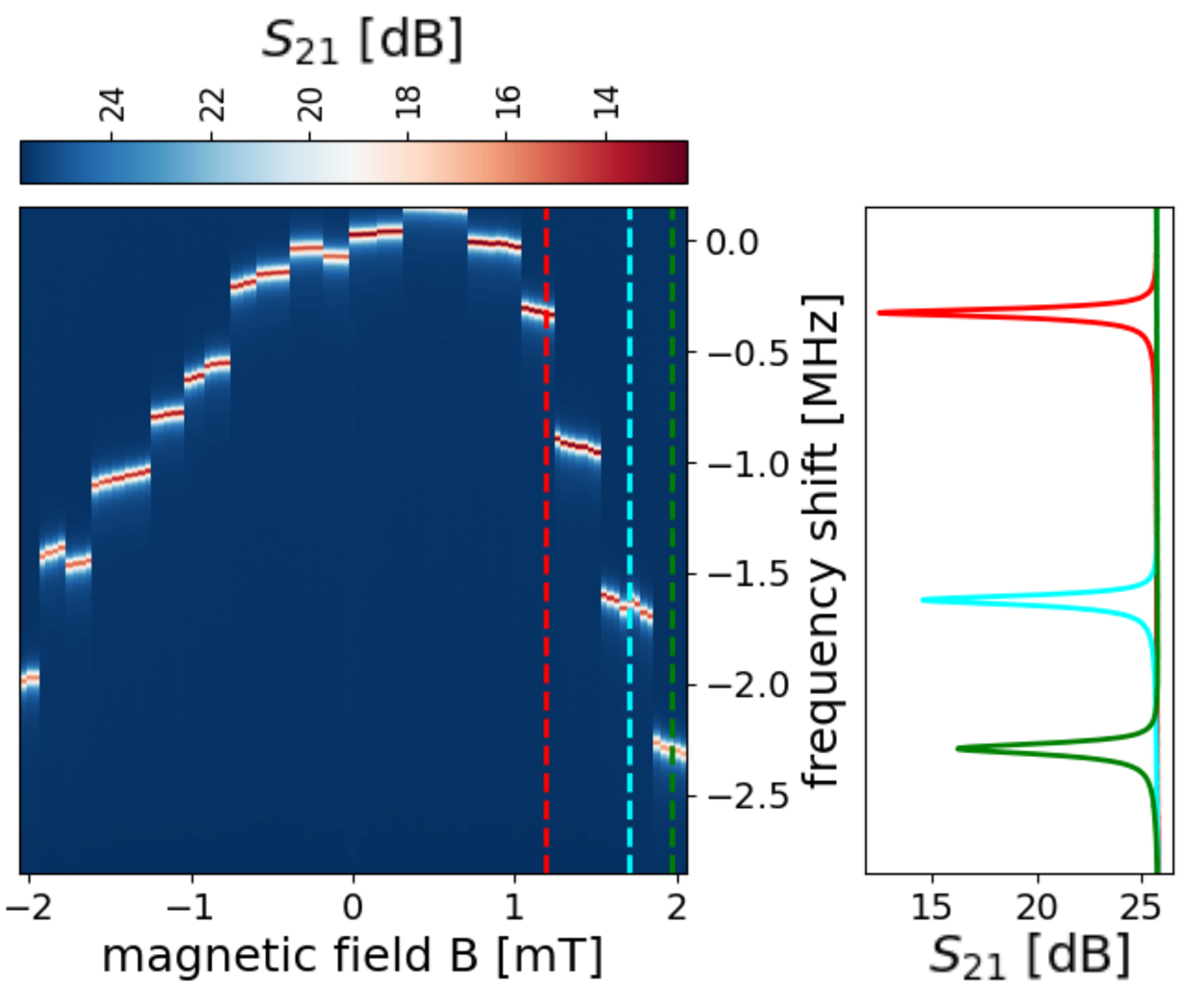}
}
\caption{(Left) Resonator frequency shift as a function of applied magnetic field current, showing discrete steps corresponding to vortex entry events during an increasing-field sweep after nominal zero-field cooling in the shielded setup. (Right) Transmission spectra taken at three values of magnetic field (vertical dashed lines), indicating a progressive decrease in the loaded quality factor $Q_L$. (neck width $W=\SI{1}{\micro\meter}$, T = 20 mK).}
\label{fig2}
\end{figure}

Magnetic-field imaging was performed in a separate optical cryostat using a quantum diamond microscope (QDM) based on a perfectly aligned nitrogen-vacancy (NV) ensemble~\cite{SN2023,Tsuji2022,SN2025}. The QDM data were acquired on a resonator with a \(W=2~\si{\micro\meter}\) neck, chosen to match the optical resolution; this device also exhibits discrete microwave steps, whereas the data in Fig.~\ref{fig2} were taken at millikelvin temperature on a \(W=1~\si{\micro\meter}\) neck.
The QDM field history was set by cooling the device to \(3~\si{\kelvin}\) in the optical cryostat, while laser heating during NV readout gives an imaging temperature of \(\sim6.4~\si{\kelvin}\)~\cite{SI}. The QDM and millikelvin microwave measurements are therefore not intended as a field-by-field correlation experiment. Their comparison is microscopic and geometric: QDM imaging tests where vortices enter in the lithographic neck geometry, while microwave spectroscopy probes how such localized entry appears in the resonator impedance. A simultaneous QDM--microwave experiment would not be a like-for-like repetition of Fig.~\ref{fig2}, because placing the diamond sensor close to the CPW would substantially dielectric-load the resonator and modify the resonance frequency, coupling, and microwave-current distribution.

\noindent\textit{Discrete Frequency Shifts from Vortex Entry.—}
We begin by analyzing the microwave response of the device with $1~\si{\micro m}$-wide constriction region. When we sweep the external magnetic field, we observe a smooth, approximately quadratic reduction in the resonance frequency of the $\lambda/4$ resonator. This behavior is attributed to the increasing London penetration depth under a magnetic field, which effectively modifies the resonator inductance~\cite{Healey2008}. Superimposed on this smooth background, we detect discrete, abrupt downward steps in the resonance frequency, typically spaced by 0.2--0.5~MHz (Fig.~\ref{fig2}). 
These features are reproducible and are consistently observed across multiple devices. 

Each frequency step is accompanied by a measurable reduction in the loaded quality factor $Q_L$, suggesting an additional dissipative channel activated at each event. 
The magnitude of the frequency shift and the quality factor drop vary from step to step (see Supplementary material~\cite{SI}), and similar step-like structures appear in both up- and down-sweeps of the magnetic field. The absolute field values of the jumps vary between scans because vortex entry is hysteretic and depends on magnetic history and microscopic pinning. The comparison with the End Matter data is therefore intended to show the dependence on constriction width, not identical entry fields.
In addition, we observe pronounced hysteresis (see End Matter): upon reversing the sweep direction, the resonance frequency does not immediately return to its original trajectory, indicating that the underlying state of the device after each step remains altered. 

To understand the dependence of these discrete features on device geometry, we compare resonators with different constriction widths. Narrower constrictions result in larger and more frequent frequency jumps, while no discernible frequency jumps are observed in devices with wider constrictions (see End Matter for details).
This constriction-width dependence also disfavors an origin in remote flux avalanches \cite{nulens2023catastrophic}. A flux instability occurring in the ground plane or in a distant part of the center conductor would be governed primarily by the global film geometry, cooling history, and disorder, and would not be expected to disappear when only the neck width is increased. In contrast, a localized vortex in the neck naturally produces a width-dependent signal because its microwave impedance contribution scales as \(Z_v \propto W^{-2}\).

Together, these observations motivate a physical interpretation. The discrete hysteretic steps with correlated changes in $Q_i$, as well as their strong dependence on the constriction geometry, can be naturally explained by localized Abrikosov-vortex penetration events and subsequent pinning in the engineered neck. In this picture, each step is assigned to a localized vortex-entry event in which a vortex becomes trapped, producing a localized change in both the reactive and dissipative components of the impedance. 
The delayed exit and remanent frequency shift, visible in~\cref{fig2} as an offset of the parabola center, are consistent with Bean-like critical-state behavior in type-II superconductors, in which irreversible vortex motion and strong pinning lead to history-dependent electromagnetic response~\cite{Bean1962}.

\begin{figure}[tbp]
\noindent\centering{
\includegraphics[width=0.5\textwidth]{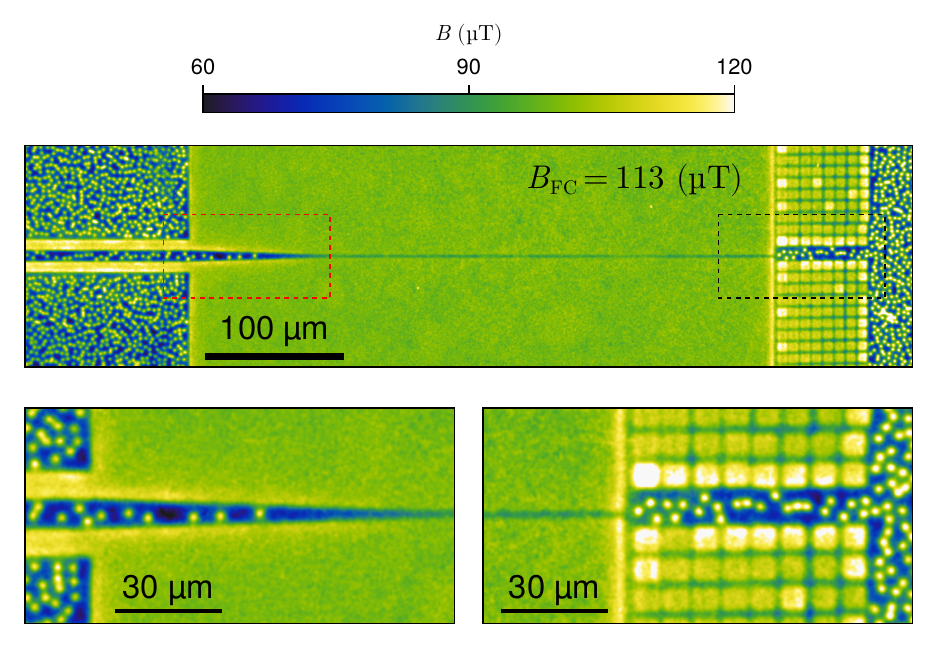}
}
\caption{(Top) QDM Magnetic field map over a wide field of view in the neck region of the device with a $2~\si{\micro\meter}$-wide neck after field cooling in $113~\si{\micro\tesla}$.
(Bottom left) Zoomed-in view of the red dashed rectangular region in the top panel.
(Bottom right) Zoomed-in view of the black dashed rectangular region in the top panel.}
\label{fig3}
\end{figure}

\begin{figure*}[tbp]
\begin{center}
\includegraphics[width=\textwidth]{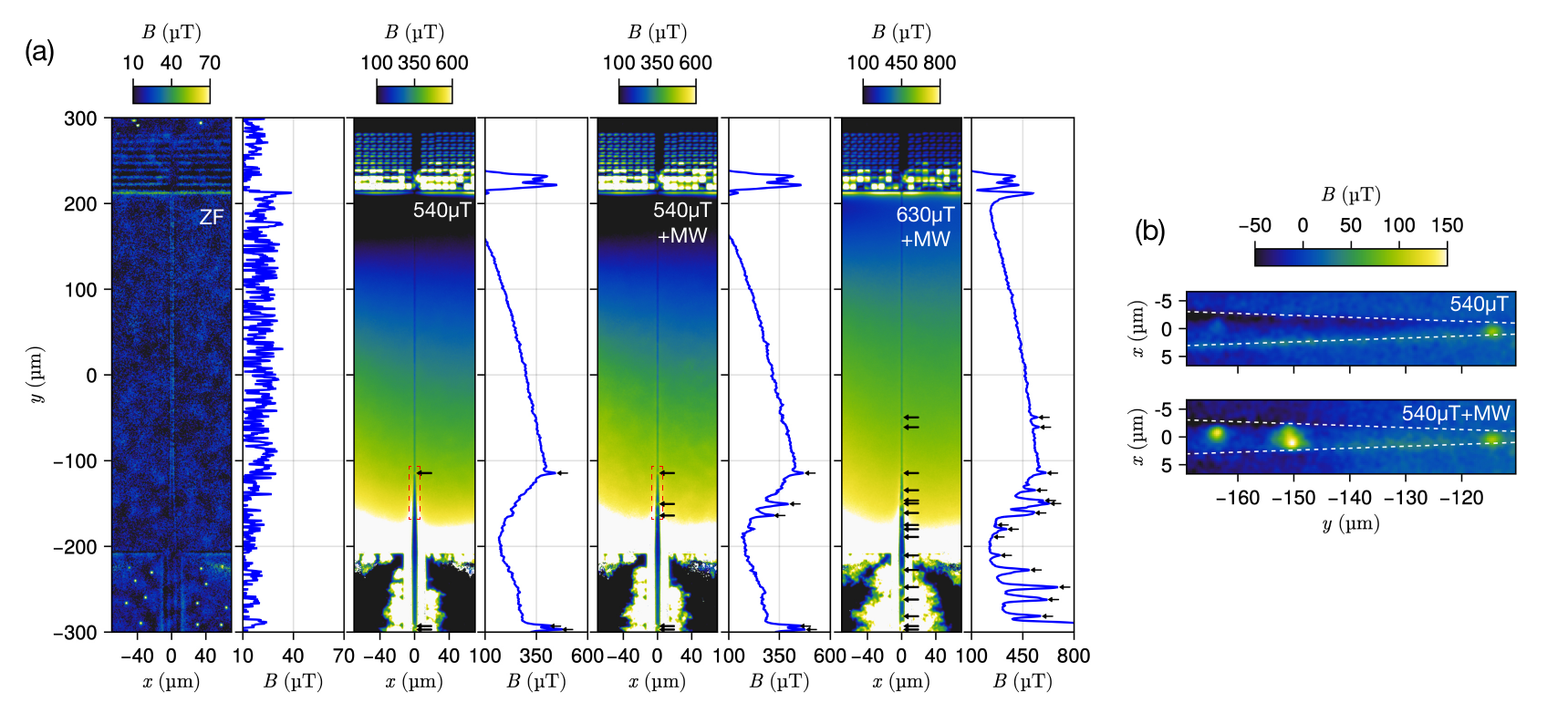}
\caption{
  (a) Magnetic-field imaging. 
  (ZF) Magnetic-field map after cooling in a near-zero field (\( \lesssim \)\SI{1}{\micro\tesla}). 
  (540~\si{\micro T}) After increasing the external field to \(\sim\)\SI{540}{\micro\tesla}. 
  (540~\si{\micro T}+MW) After sweeping the microwave (MW) drive across resonance. 
  (630~\si{\micro T}+MW) After increasing the field to \(\sim\)\SI{630}{\micro\tesla} and repeating the MW sweep. 
  (b) Enlarged views of the red dashed region in (a) for (540~\si{\micro T}) and (540~\si{\micro T}+MW) (Red dashed region). The slow background gradient visible in (a) is removed by subtracting a reference magnetic image acquired at 450~\si{\micro T}.
  Vortices are located near the edge. (neck width $W=\SI{2}{\micro\meter}$, imaging at $\sim6.4$ K).
}
\label{fig4}
\end{center}
\end{figure*}

\noindent\textit{Magnetic-Field Measurements Using NV Centers.-}
To independently support the assignment of the discrete microwave features to localized vortex-entry events, we directly image the magnetic flux at the device surface.
The QDM measurements were performed under the separate experimental conditions summarized in the Materials and Methods section.

To validate our magnetic imaging technique, we first perform FC on the device with a $2~\si{\micro\meter}$-wide neck.
\cref{fig3} shows a magnetic-field image of the region around the neck after FC to \SI{3}{K} in \SI{113}{\micro\tesla}. 
Individual vortices are clearly visible both in the Nb ground planes of the CPW and in the wider section of the center conductor.
However, no vortices are observed in the neck region.
The absence of trapped vortices in the neck after FC should be interpreted as a narrow-strip vortex-expulsion effect rather than as a simple enhancement of the bulk lower critical field. For a thin-film strip, the field-cooled trapping threshold scales approximately as \(B_{\rm trap}\sim\Phi_0/W^2\), and for the present \(W=2~\si{\micro\meter}\) QDM device the estimate \(B_{\rm K}\simeq1.65\Phi_0/W^2\) gives \(B_{\rm K}\approx0.85~\si{\milli\tesla}\), well above the \(113~\si{\micro\tesla}\) field used here~\cite{Stan2004,Kuit2008}.
This FC trapping scale should be distinguished from the first-entry field during an increasing-field sweep after ZFC. The latter is controlled by Bean--Livingston/edge and geometrical barriers, lithographic edge disorder, current crowding, and the nonlocal Meissner-current distribution of the washer geometry. A detailed discussion, including wide-field QDM images showing the suppressed-field band in the washer/window, is given in the Supplemental Material~\cite{SI}. 
\cref{fig4}(a) shows a magnetic-field image of the region around the neck after near-zero-field cooling (ZFC) to \SI{3}{K} in $\lesssim\SI{1}{\micro\tesla}$.
The far-left panel (labelled ZF) shows the magnetic field right after cooldown. 
While some individual vortices still appear in the Nb ground planes of the CPW, no vortices are observed in the whole center conductor.
The right panel shows the field profile averaged over a $\pm\SI{1}{\micro\meter}$-wide strip centered on the neck axis, which exhibits a noisy background due to measurement noise.

We start from this pure Meissner state after ZFC and, while keeping $T<T_c$, apply additional magnetic field and MW drive.
Upon increasing the field to $\sim\SI{540}{\micro\tesla}$ (labeled \SI{540}{\micro\tesla}), flux begins to penetrate the Nb.
At these fields, the large-field-of-view QDM images show localized trapped flux, remote edge-flux penetration, and a smooth global background associated with the washer geometry. Importantly, we do not observe an avalanche-like flux front crossing the neck/window or the high-current center-conductor region where the discrete vortices are resolved.

We then sweep the microwave (MW) drive across resonance from outside the cryostat, thereby exciting MW currents in the resonator can assist vortex entry by periodically driving the edge current toward the critical value.
The subsequent image (labelled \SI{540}{\micro\tesla} + MW) shows two additional vortices near the thinnest section of the neck.
After further increasing the field to $\sim\SI{630}{\micro\tesla}$ and repeating the MW sweep, multiple vortices enter the neck region, producing an oscillatory signature in the $x$-profile of the magnetic field.

In the magnetic images we observe that, under both magnetic and microwave drive, vortices nucleate as expected.
Furthermore, vortices preferentially sit near the taper region between the CPW and the thin microstrip.
Within the flux-focusing window (the Nb-etched region), the magnetic field exhibits pronounced spatial gradients: near the top of the opening the field is strongly suppressed, whereas near the bottom it is significantly enhanced and focused.
This spatial variation of the magnetic field is linked to the neck geometry. As the strip narrows, the current redistributes across its width and becomes stronger near the edges than in the center. The lithographic shape of the constriction is shown in~\cref{fig4}(b).
In this region, the microstrip geometry varies along the longitudinal coordinate and individual vortices occupy different longitudinal positions, which may account for the varying step sizes observed in the discrete resonance-frequency shifts.

This inhomogeneity explains the tendency for vortices to nucleate in the taper region.
When averaged over the entire flux-focusing window area, the magnetic field is suppressed, rather than enhanced, relative to the externally applied field.
This behavior is consistent with the geometrical effect of a closed flux-focusing window described by \citet{Brandt2005}, in which the current stream function $g(r)$ is determined by the superconductor's global geometry, and the magnetic field distribution inside the annulus is determined by $g(r)$.
	This spatially inhomogeneous field profile also explains why the narrowest part of the constriction can remain vortex-free after cooldown even when vortices are visible in wider regions: the relevant quantity for vortex entry is not the externally applied field alone, but the local perpendicular induction produced by the applied field together with the nonlocal Meissner screening currents of the patterned film.

The QDM and millikelvin microwave measurements were performed under different experimental conditions, so the absolute vortex-entry fields are not expected to coincide quantitatively. The relevant comparison is geometric and microscopic: QDM imaging directly shows that vortices enter the lithographic neck discretely, while the microwave data show discrete, hysteretic impedance changes strongly enhanced by narrow constrictions.

\noindent\textit{Discussion and Conclusions.—}
We have shown that narrow lithographic necks integrated into thin-film Nb $\lambda/4$ CPW resonators enable \emph{real-time} observation of microwave signatures of individual Abrikosov-vortex penetration events as discrete, step-like downward shifts of the resonance frequency $f_0$ accompanied by concomitant increases in $1/Q_i$.
The step amplitudes and event counts correlate with the neck geometry, and the observed hysteresis between up- and down-sweeps is consistent with strong pinning at the engineered constriction.

The sensitivity of the resonator response to the constriction width can be understood as follows. The impact of a single vortex in the constriction is that of an impedance added in series to the resonator circuit. Using expressions for the resistivity $\tilde{\rho}_{ff}(\omega)$ due to a finite vortex density $\frac{B}{\Phi_0}$ \cite{pompeo2008reliable,Song2009PRB}, one can derive (see Supplementary material~\cite{SI}) the impedance of a single vortex in a constriction of width $W$ and thickness $t$: $Z_v(\omega) = \tilde{\rho}_{ff}(\omega)/B*\Phi_0/(W^2 t) \propto W^{-2}$. As a result, $Z_v$ becomes significantly larger at lower $W$, consistent with experimental observations.

A related recent study of Nb CPW resonators reported staircase-like resonance jumps under perpendicular-field sweeps and attributed them to Abrikosov-vortex entry and exit, but in a geometry without the same neck localization and magnetic imaging those jumps were interpreted as multiple-vortex events~\cite{Kalashnikov2026Equidistant}. This comparison underscores that our assignment relies not on step-like microwave data alone, but on the engineered neck-width dependence and QDM observation of discrete vortex entry at the constriction.

Unlike scanning-probe techniques, our VNA readout is fast, multiplexable, and free of moving parts. 
In contrast to dc-transport signatures in nanowires and Josephson devices~\cite{mills2016vortex,ligato2021preliminary}, our dispersive readout does not require bias currents and minimizes self-heating,  while preserving sensitivity to single-vortex events. The ability to resolve discrete dispersive shifts suggests that individual vortices can, in principle, function as localized, reconfigurable information carriers within superconducting microwave circuits. The key advantage of our approach with regard to previous ones \cite{kalashnikov2024demonstration, Foltyn2024} is the combination of low losses, which allows single-vortex states to be distinguished with high fidelity and strong coupling to microwave currents. This level of visibility, together with the compatibility of resonator architectures with multiplexed cryogenic electronics, indicates that field-controlled access to metastable vortex configurations could form the basis of a compact and scalable memory platform.

Looking forward, the same platform can enable controlled studies of stochastic entry and depinning in engineered pinning potentials (e.g., necks, notches, or antidots) placed at current antinodes, two-tone spectroscopy to probe dynamics near the vortex depinning frequency $f_d$, angle-dependent field studies, and large-scale multiplexed arrays for statistics over nominally identical geometries.
Beyond basic vortex physics, immediate applications include diagnosing and mitigating vortex-induced loss in superconducting microwave circuits (qubits and resonators) and realizing simple on-chip single-vortex metrology compatible with cryogenic integration.

\textit{Acknowledgments.—}
The authors thank Yusuke Kato for insightful discussions. This work was partially supported by JST, CREST Grant No. JPMJCR23I2, Japan; Grants-in-Aid for Scientific Research (Nos. JP25H01248, JP24K21194, 23K25800,  JP22K03524,  JP25K00934,); Seiko Instruments Advanced Technology Foundation Research Grants;
the Cooperative Research Project of RIEC, Tohoku University; ``Advanced Research Infrastructure for Materials and Nanotechnology in
Japan (ARIM)'' (No. JPMXP1222UT1131) of the Ministry of Education, Culture,
Sports, Science and Technology of Japan (MEXT); the MEXT Quantum Leap Flagship Program (MEXT Q-LEAP) Grant Number JPMXS0118067395;
ARIM Grant Numbers  JPMXP1224UT1047 and JPMXP1225UT1027 and supported by JSPS KAKENHI Grant Number 23H04864.
This work was carried out using the joint-use facilities of the Millikelvin Quantum Platform at the Cryogenic Research Center, The University of Tokyo.

\textit{Author Contributions.—}
KS conceived the idea, coordinated the project, and performed the microwave resonator experiments.
SN performed vortex imaging experiments using QDM under the supervision of KeS and KK. 
RH performed early-stage experiments for QDM imaging. KeS and KK provided the QDM imaging setup.
PAV developed the theoretical framework and contributed to data interpretation.
MH and TI fabricated the superconducting samples.
TT, TIw, and MHa provided the diamond device with NV centers and experimental infrastructure.
All authors contributed to manuscript preparation.

\bibliography{main}

\clearpage
\appendix
\newpage

\section{\Large End Matter}

\textit{Theoretical model.—}
\label{sec:theory}
We model the effect of a single Abrikosov vortex, pinned in the engineered neck of the Nb coplanar waveguide ($\lambda/4$) resonator, as a small, frequency-dependent series impedance $Z_v(\omega)$ inserted at a current antinode. This perturbation shifts the resonance frequency $f_0$ and the internal quality factor $Q_i$. The description combines driven-vortex electrodynamics with a local participation model that captures current crowding in the constriction.

The impedance induced by a single vortex can be obtained from the model expression for complex vortex resistivity \cite{pompeo2008reliable,Song2009PRB} $\tilde{\rho}_{ff}(\omega) \;=\; \frac{\Phi_0 B} {\eta}\frac{\epsilon + i\,\omega/\omega_d}{1 + i\,\omega/\omega_d}$, which assumes an areal density of vortices $n_v = \frac{B}{\Phi_0}$, $\Phi_0$ being the superconducting flux quantum. The complex impedance produced by a single vortex is then obtained as $Z_v(\omega) = \tilde{\rho}_{ff}(\omega)*l/(W t)/(n_v*l*W)$, where $W,l,t$ are the width, length and thickness of the superconducting region, respectively. In addition, since the current can be distributed inhomogenously along the width of the constriction, the current that acts on the vortex can be larger or smaller than the average current. We account for this effect by a factor $g_I$. The result reads
\begin{equation}
Z_v(\omega) \;=\; g_I \frac{\Phi_0^2} {\eta t W^2}\frac{\epsilon + i\,\omega/\omega_d}{1 + i\,\omega/\omega_d}
\label{eq:Zv_local}
\end{equation}
One can estimate $\Phi_0/\eta$ in Nb from resistivity in the flux-flow regime, where $\rho_{ff} = \frac{\Phi_0 B}{\eta}$. In \cite{Nb_ff}, $\rho_{ff} \sim 10^{-9} \Omega\cdot m$ at $B=30 mT$, so we get $\frac{\Phi_0}{\eta} \approx 3 \cdot 10^{-8} \Omega\cdot m/T$. Note however that the measurements in \cite{Nb_ff} were done at T=7.8 K.

Generally, the observable fractional changes in frequency and quality factor of the cavity can be related to the impedance change as \cite{ybco_imped,Song2009}:
\begin{equation}
    \frac{\delta f}{f_0} \;\simeq\; -\,\frac{X_v(\omega_0)}{2 G},
\qquad
\delta\!\left(\frac{1}{Q_i}\right) \;\simeq\; \frac{R_v(\omega_0)}{G}\,,
\label{eq:df_dQi_master}
\end{equation}
where $Z_v = R_v+i X_v$ and $\delta f = f-f_0$ (hence the sign change with respect to \cite{Song2009}) and $G$ is a parameter with dimensions of impedance that can depend on the geometric details as well as the current distribution and the kinetic inductance contribution to the resonator impedance~\cite{Song2009}.

\textit{Constriction-width dependence of the vortex-induced response.—}
\label{sec:neck-width-suppl}
In the main text we focus on the device with a $1~\mu$m-wide neck, where individual AVs entering
the constriction produce clearly resolved, step-like shifts of the resonance frequency $f_0(B)$ and correlated changes in
the internal quality factor $Q_i$. Here we provide additional data and context for the dependence of this response on the
neck width.

\begin{figure}[h!]
  \centering
  \includegraphics[width=0.95\columnwidth]{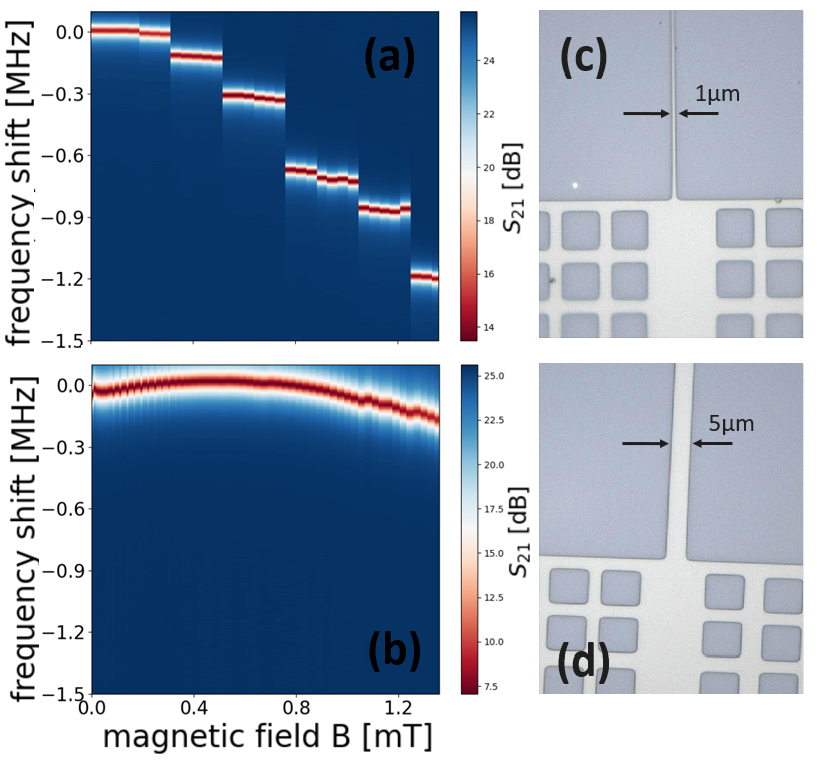}
  \caption{
    \textbf{Constriction-width dependence.}
    (a) Magnetic-field dependence of the resonance frequency for the resonator with a $1~\mu$m-wide neck, measured in an increasing-field scan on the same nominal constriction geometry as Fig.~\ref{fig2} but in a different field scan, showing
    multiple discrete steps attributed to individual vortex entry events.
    (b) Same measurement for the $5~\mu$m-wide neck, taken on a different wider-neck device/sample, where the response is smooth within our resolution.
    (c,d) Optical micrographs of the corresponding neck geometries.
    Narrower constrictions amplify the single-vortex impedance and the local current density, yielding larger and more
    frequent resolvable steps in $f_0(B)$. The comparison is intended to show the constriction-width dependence, not identical vortex-entry fields, because vortex entry is hysteretic and depends on magnetic history and microscopic pinning.
  }
  \label{fig:neck-width-suppl}
\end{figure}

Fig.~\ref{fig:neck-width-suppl} compares the magnetic-field dependence of the resonance frequency for two resonators
in different devices/samples: one with a $1~\mu$m-wide neck and one with a $5~\mu$m-wide neck, together with optical micrographs
of the corresponding geometries. For the $1~\mu$m device, the $f_0(B)$ trace exhibits multiple discrete, downward
frequency steps as the out-of-plane field is increased, consistent with the entry and pinning of individual AVs in the
constriction region. In contrast, the $5~\mu$m device shows a smooth, approximately quadratic background shift with no
discernible discrete jumps within our resolution; any vortex-related contribution is effectively averaged out over the
wider cross-section. Because the traces were acquired in different scans/devices, the absolute entry fields should not be compared directly; the robust comparison is the presence or absence of resolvable steps as the neck width is changed.

This neck-width dependence provides an additional argument against a remote-avalanche origin. Flux motion in the ground plane or in a distant part of the center conductor would not naturally be switched on or off by changing only the lithographic neck width, whereas a vortex localized in the neck is expected to produce a strongly width-dependent microwave response because $Z_{\mathrm{v}}\propto W^{-2}$.

This trend is naturally understood within the single-vortex impedance model.
There we model a vortex pinned in the neck as a small series impedance
$Z_{\mathrm{v}}(\omega)$ inserted at a current antinode, and obtain
\begin{equation}
  Z_{\mathrm{v}}(\omega) \propto \frac{1}{W^2 t},
\end{equation}
where $W$ and $t$ are the width and thickness of the superconducting strip, respectively. The $W^{-2}$ scaling implies that
the reactive and dissipative perturbations produced by a single vortex grow rapidly as the constriction is narrowed.
Combined with current crowding in the neck, this enhancement makes the step height $\Delta f_0$ per vortex in the
$1~\mu$m device large enough to be resolved as discrete events, whereas in the $5~\mu$m neck the per-vortex shift is
comparable to, or smaller than, the smooth background trend and experimental noise.

From the point of view of mode participation, the total energy stored in the neck region is small compared with the full resonator length, but the local kinetic-energy density is strongly enhanced by the geometric current concentration.

\begin{figure}[h!]
  \centering
  \includegraphics[width=0.95\columnwidth]{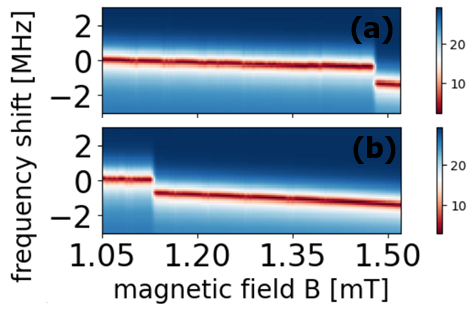}
  \caption{
    \textbf{Magnetic hysteresis of a single vortex entry/exit event.}
    Color-scale map of the transmission amplitude $|S_{21}|$ as a function of frequency detuning and coil current for magnetic-field sweeps (a) increasing and (b) decreasing. The step-like shift of the resonance occurs at different magnetic-field values depending on the sweep direction, demonstrating hysteretic behavior associated with vortex pinning in the constriction. This measurement was taken on the nominal $W=1~\textmu m$ neck device in a field cycle distinct from the increasing-field scan shown in Fig.~\ref{fig2}; therefore the absolute jump fields need not coincide.}
  \label{fig:hysteresis}
\end{figure}

\textit{Magnetic Hysteresis.—}
The entry of vortices in a superconductor is known to be a hysteretic process~\cite{Tinkham1996IntroSuperconductivity, Zeldov1994Geometrical}. Once a vortex enters a superconducting region driven by increasing magnetic field, reducing the field below the vortex entry value is typically not sufficient to expel the vortex. As a consequence, the magnetic field at which a given frequency step appears may be expected to depend on the magnetic history of the device. This is directly demonstrated in Fig.~\ref{fig:hysteresis}, where the same step-like resonance shift occurs at different field values for increasing and decreasing magnetic-field sweeps, providing clear evidence of hysteretic behavior associated with vortex entry and exit. For this experiment, we used a Nb coplanar superconducting resonator with a fundamental frequency of 11.77~GHz. The resonator includes a narrow neck section with a linewidth of 1~\textmu m.

This hysteresis is naturally explained by vortex pinning in the narrow neck of the resonator. Once a vortex becomes trapped at a pinning site, an additional magnetic-field bias is required to depin it and drive its motion out of the constriction. The need for different effective fields during vortex entry and exit reflects the finite energy barrier associated with vortex-pinning interaction, consistent with previous observations of hysteresis in superconducting coplanar resonators~\cite{Bothner}. Within this picture, the observed hysteresis arises from irreversible vortex dynamics rather than from changes in the intrinsic resonator properties. This hysteresis shows that different vortex configurations in the constriction can be accessed reproducibly by magnetic-field sweeps, consistent with strong pinning and delayed vortex exit.

\medskip
\medskip
\medskip
\medskip
\medskip

\begin{center}
\small
\textcopyright{} 2026 American Physical Society. 
This is the accepted manuscript of the following article:

\medskip

K.~Shulga, et al. ``Observation of Individual Vortex Penetration in a Coplanar Superconducting Resonator,'' 
\textit{Phys. Rev. Lett.} \textbf{137}, 116001 (2026).

\medskip

The final published version is available at 
\href{https://doi.org/10.1103/4zw3-hm6t}
{https://doi.org/10.1103/4zw3-hm6t}. 
This manuscript version is posted with permission for 
non-commercial scholarly use.
\end{center}

\clearpage
\appendix
\newpage

\section{\Large Supplemental Material}

\section*{Coil calibration}
The calibration based on geometric coil parameters yielded a conversion factor of 0.034~mT/mA, with accuracy limited by Meissner screening. The sweep range (typically $\pm$60~mA) was covered in 5–10 minutes, allowing thermally activated vortex creep and stable pinning. The applied magnetic fields in our experiments remained in the low-field regime. The maximum coil current of 60~mA corresponds to an estimated field of approximately 2~mT at the chip center. This is well below the upper critical field $H_{c2}$ of Nb thin films (on the order of several tesla), but exceeds the lower critical field $H_{c1}$, which is suppressed in thin-film geometries and typically falls below 1~mT~\cite{Stan2004}. Thus, the observed vortex entry occurs under conditions where individual flux quanta are expected to penetrate the film.

\section*{Microwave readout and wiring}
Microwave readout was performed using a vector network analyzer in continuous-wave mode, capturing both amplitude and phase. The input line was attenuated by 70~dB (distributed over temperature stages), while the output signal passed through two circulators and a 4–8~GHz bandpass filter before being amplified by a 4~K HEMT amplifier and a low-noise room-temperature chain (total gain ~60~dB). The frequency sweep resolution was sufficient to resolve sub-MHz resonance shifts.

\section*{Vortex trapping, field redistribution and entry barriers at the neck}

\subsection*{Field-cooled trapping scale and ZFC first-entry field}

The absence of trapped vortices in the narrow neck after field cooling and the first vortex-entry field during an increasing-field sweep after zero-field cooling involve related but distinct field scales.
After field cooling, the relevant question is whether a vortex remains trapped in a thin-film strip of width \(W\).
Previous work on superconducting strips showed that the threshold field for complete vortex expulsion scales approximately as
\[
    B_{\rm trap}\sim \frac{\Phi_0}{W^2}.
\]
More specifically, Kuit et al.~\cite{Kuit2008} modeled the Gibbs free energy of a vortex in a thin-film strip, including the vortex self-energy, the Meissner energy, and vortex--antivortex-pair formation, and found the onset of trapping to be
\[
    B_{\rm K}\simeq 1.65\frac{\Phi_0}{W^2}.
\]
For the \(W=2~\si{\micro\meter}\) neck used in the QDM images, this gives
\[
    B_{\rm K}\approx 0.85~\si{\milli\tesla},
\]
which is well above the \(113~\si{\micro\tesla}\) field used for the field-cooled image in Fig.~\ref{fig3}.
The vortex-free neck after field cooling is therefore consistent with the narrow-strip vortex-expulsion scale~\cite{Stan2004,Kuit2008}.

This argument should not be confused with the first-entry field \(H_p\) during an increasing-field sweep after ZFC.
The latter is governed by surface and edge barriers, geometrical barriers, lithographic edge disorder, current crowding, and the nonlocal Meissner-current distribution of the patterned washer geometry, rather than by a simple \(\Phi_0/W^2\) law.

\subsection*{Bean--Livingston and geometrical barriers}

There are two representative barriers that delay vortex entry in a type-II superconductor.
The Bean--Livingston (BL) surface barrier originates from the competition between the attraction of a vortex to its image antivortex near the surface and the Lorentz force from the Meissner screening current, which pushes the vortex into the superconductor~\cite{BeanLivingston1964}.
For an ideally smooth surface, vortices need not enter immediately when the applied field exceeds \(H_{c1}\); instead, the metastable Meissner state can persist until the surface current approaches the instability scale associated with the thermodynamic critical field, \(H_c\sim \kappa H_{c1}/\ln\kappa\), where \(\kappa=\lambda/\xi\).

The geometrical barrier has a different origin.
It arises from the nonuniform Meissner-current distribution and the position dependence of the vortex self-energy in a finite platelet or strip in perpendicular field~\cite{Zeldov1994Geometrical}.
For an ideal rectangular strip of width \(2w\) and thickness \(d\), Brandt, Mikitik, and Zeldov analyzed the competition between the BL and geometrical barriers~\cite{Brandt2013}.
In an order-of-magnitude geometrical-barrier estimate,
\[
    H_p^{\rm GB}
    \simeq H_{c1}\sqrt{m}\cos\theta,
    \qquad
    m\simeq \frac{2d}{\pi w},
    \qquad
    \cos\theta\simeq 0.80,
\]
so that
\[
    H_p^{\rm GB}\sim H_{c1}\sqrt{\frac{d}{w}}.
\]
The BL-limited penetration field for vortex nucleation at the strip corner is estimated as
\[
    H_p^{\rm BL}
    \simeq
    \frac{0.92 H_{c1}\kappa}{\ln\kappa}
    \left(
        \frac{18d\lambda^2}{\pi w^3}
    \right)^{1/6}.
\]
The crossover between the two regimes is controlled by
\[
    p =
    \frac{\kappa}{\ln\kappa}
    \left(\frac{\lambda}{d}\right)^{1/3},
    \qquad
    p_c\simeq 0.52.
\]
For \(p>p_c\), first penetration is determined by the disappearance of the BL barrier at the strip corner, whereas for \(p<p_c\), geometrical-barrier physics gives a two-stage penetration process~\cite{Brandt2013}.

Using representative Nb thin-film parameters,
\[
    \lambda\simeq 92~\si{\nano\meter},
    \qquad
    \xi\simeq 12~\si{\nano\meter},
    \qquad
    d\simeq 200~\si{\nano\meter},
\]
one obtains
\[
    \kappa=\frac{\lambda}{\xi}\simeq 7.7,
    \qquad
    p\simeq
    \frac{\kappa}{\ln\kappa}
    \left(\frac{\lambda}{d}\right)^{1/3}
    \simeq 2.9 \gg p_c.
\]
Thus, the ideal-strip estimate places the present Nb thin-film neck in the BL/edge-barrier-dominated regime, consistent with the discussion in Ref.~\cite{Silhanek2025}.
Even if one uses a more conservative coherence length, e.g. \(\xi\simeq40~\si{\nano\meter}\), one still obtains \(p>p_c\).

These ideal-strip formulas should be viewed only as qualitative guidance for the present lithographically patterned resonator.
In an actual Nb film, edge roughness, etch damage, local oxidation, tapering of the constriction, border defects, and current crowding can locally reduce the penetration barrier and select the vortex nucleation site~\cite{Silhanek2025}.
Therefore, the experimentally observed first-entry field is not expected to be given quantitatively by the ideal-strip expressions.
The important point for the present work is instead that vortex entry into the neck is controlled by the combined effect of local edge barriers and the nonlocal magnetic-field profile of the full washer geometry.

\subsection*{Meissner-field redistribution in the washer geometry}

The neck is not an isolated strip but part of a multiply connected patterned Nb film containing a washer/window.
In such a geometry, the local perpendicular induction near the neck is not simply equal to the externally applied field \(B_a\).
Instead, it is determined nonlocally by the Meissner screening currents flowing over the full superconducting pattern and by the flux-focusing washer/window geometry.

A convenient way to describe this effect is the thin-film stream-function formulation~\cite{Brandt2005}.
The sheet-current density can be written as
\[
    \mathbf{J}(x,y)=\hat{\mathbf{z}}\times\nabla g(x,y),
\]
where \(g(x,y)\) is the stream function.
In a multiply connected film, the total stream function contains both a component that screens the applied field and a component associated with circulating current around the washer:
\[
    g_{\rm tot}=B_a \tilde g_1 + I\tilde g_2.
\]
Here, \(\tilde g_1\) is a global shielding-current component determined by the full film shape, while \(\tilde g_2\) represents the current associated with trapped flux or circulating current around the hole/window.
Because \(\tilde g_1\) is nonlocal, the field distribution inside the washer can be strongly inhomogeneous, even when the externally applied field is uniform.

This global Meissner-screening pattern can create a zero-field or suppressed-field band inside the washer/window.
As \(B_a\) is increased, the total flux entering the washer changes, while the suppressed-field band can persist and move through the structure.
Consequently, the local perpendicular induction near the narrow stripline can remain much smaller than the externally applied field.
This provides an additional reason why vortex entry into the neck is delayed and occurs only at selected positions: the local edge barrier is high, and the local magnetic field driving vortex entry is itself reduced by the global washer geometry.

\begin{figure*}[tbp]
    \centering
    \includegraphics[width=0.95\linewidth]{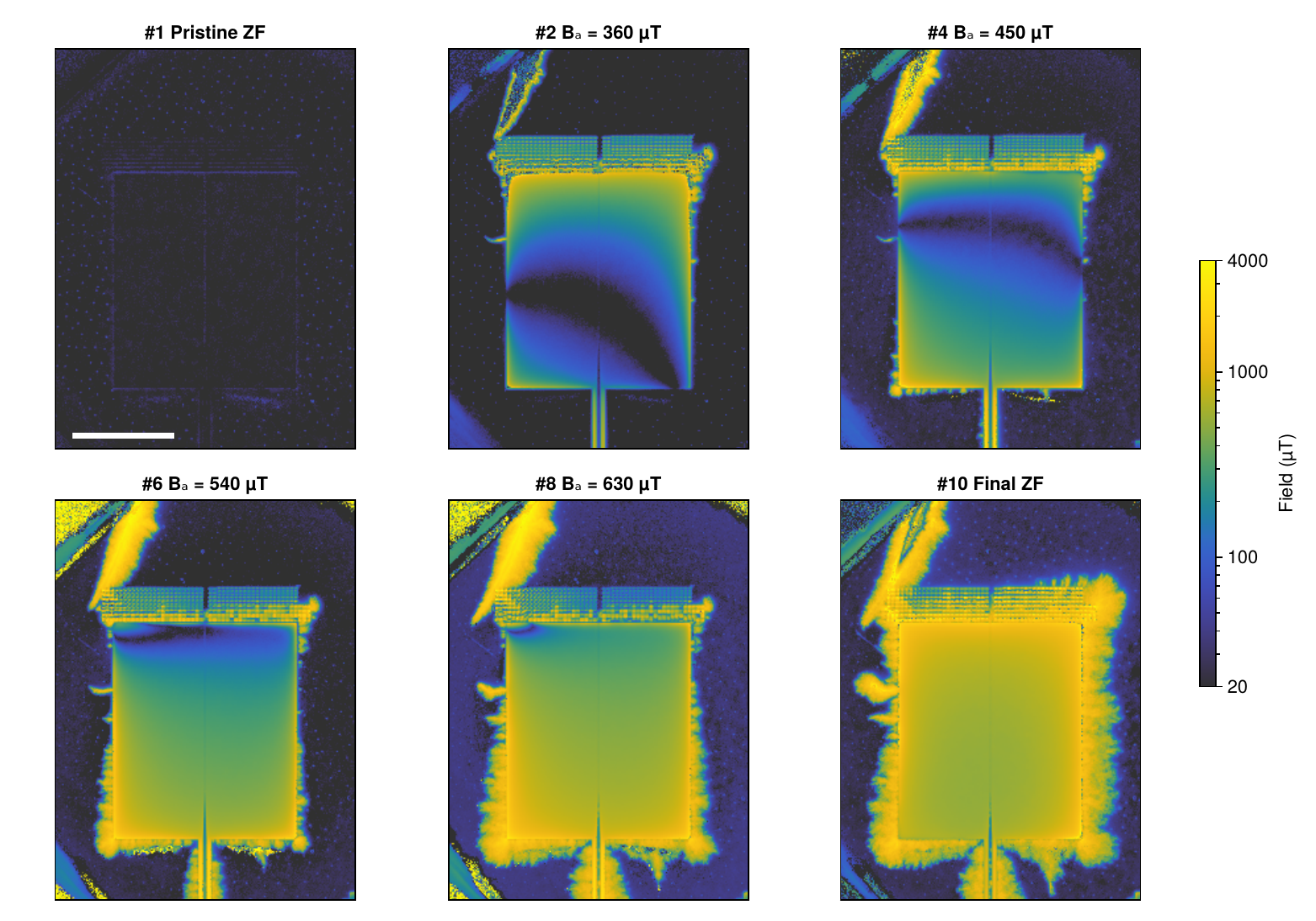}
    \caption{
    \textcolor{black}{Wide-field QDM magnetic-field images acquired while increasing the external magnetic field after ZFC.
    The panels show pristine ZF, \(B_a=360\), 450, 540, and \(630~\si{\micro\tesla}\), and final ZF.
    A suppressed-field band forms inside the washer/window and moves as the external field is increased, demonstrating that the local perpendicular induction near the narrow stripline is not simply equal to the externally applied field.
    Irregular flux penetration and trapped flux are visible at remote edges, but these images are not used to claim that thermomagnetic avalanches are impossible everywhere on the chip.
    Rather, they show that the neck/window region is governed by a smooth global Meissner-field redistribution and that no avalanche-like flux front is observed to cross the neck/window or the high-current center-conductor region in the field range where discrete neck vortices are resolved.}
    }
    \label{fig:wide_nv_zfc}
\end{figure*}

Figure~\ref{fig:wide_nv_zfc} shows this redistribution directly.
In the pristine ZF state, no vortices are observed near the center conductor.
As \(B_a\) is increased to \(360\)--\(630~\si{\micro\tesla}\), a strongly nonuniform magnetic-field distribution develops inside the washer/window; in particular, a dark suppressed-field band crosses the washer and moves through it.
These wide-field QDM images therefore show that the local magnetic field near the narrow stripline is strongly redistributed by global Meissner screening and the washer/window geometry.
Together with the local edge-barrier physics discussed above, this explains why the neck can remain vortex-free after cooldown and why subsequent vortex entry occurs only at selected positions.

The same images also clarify how we use the QDM data with respect to possible avalanche scenarios.
We do not exclude flux avalanches or irregular edge penetration everywhere in the Nb film.
Indeed, remote edge-flux penetration is visible in the wide-field images.
The relevant observation is more restricted: in the field range where discrete vortices are resolved near the neck, the QDM images do not show an avalanche-like flux front crossing the neck/window or the high-current center-conductor region.
This makes a remote-avalanche origin less natural as the primary explanation of the resolved microwave steps, especially when combined with the strong neck-width dependence of the microwave response.

\subsection*{\textcolor{black}{Microwave-assisted barrier lowering}}

We next estimate the scale on which the resonant microwave current can tilt the local edge barrier and assist vortex entry at the neck.
Loaded quality factor \(Q_L = 10^5\) suggests that the on-resonance current can indeed be large enough to lower the vortex-entry barrier.
In this work we report the \emph{loaded} quality factor \(Q_L\) extracted from transmission cuts.
In general,
\[
    Q_L^{-1}=Q_i^{-1}+Q_c^{-1}.
\]
Representative complex \(S_{21}(f)\) measurements analyzed with standard notch fits confirm operation in the weakly coupled regime (\(Q_c \gg Q_i\)), so that the field dependence of \(Q_L\) primarily reflects internal dissipation rather than coupling variations.

Using the energy in a resonator
\[
    U \simeq \frac{P_{\mathrm{in}} Q_L}{\omega}
\]
and the antinode current
\[
    I_{\mathrm{pk}} \simeq \sqrt{\frac{2U}{L_{\mathrm{tot}}}},
\]
with \(\omega=2\pi f_0\), \(f_0 = 4.4\,\mathrm{GHz}\), \(Q_L = 10^5\), and a reasonable bracket \(L_{\mathrm{tot}} \sim 2\text{--}5\,\mathrm{nH}\), we obtain
\[
    I_{\mathrm{pk}} \approx 38\text{--}60\,\si{\micro\ampere}
\]
for \(P_{\mathrm{in}}=-90\,\mathrm{dBm}\) \((1\,\mathrm{pW})\), and
\[
    I_{\mathrm{pk}} \approx 380\text{--}600\,\si{\micro\ampere}
\]
for \(P_{\mathrm{in}}=-70\,\mathrm{dBm}\) \((0.1\,\mathrm{nW})\).
In the \(w=1\,\si{\micro\meter}\) neck this current produces an edge-equivalent field
\[
    \Delta B_{\mathrm{eq}}
    \simeq
    \frac{\mu_0 I_{\mathrm{pk}}}{\pi w}
    \approx
    0.4\,I_{\mathrm{pk}}[\mathrm{A}]~\mathrm{T},
\]
so numerically
\[
    \Delta B_{\mathrm{eq}} \approx 15\text{--}24~\si{\micro\tesla}
\]
for \(-90\,\mathrm{dBm}\), and
\[
    \Delta B_{\mathrm{eq}} \approx 0.15\text{--}0.24~\si{\milli\tesla}
\]
for \(-70\,\mathrm{dBm}\).

As discussed above, the factor \(\Phi_0/W^2\) should not be used as a quantitative law for the edge barrier or for the ZFC first-entry field in the present lithographic geometry.
It is only an order-of-magnitude scale associated with vortex trapping/expulsion in narrow strips after field cooling.
The actual ZFC entry threshold is set by BL/edge-barrier physics, local edge imperfections, current crowding, and the nonlocal washer-field profile.

The current lowers the barrier because the transport current creates a Lorentz force on an incipient edge vortex,
\[
    f_L=\Phi_0 K,
\]
with sheet current
\[
    K \simeq \frac{I_{\mathrm{pk}}}{w}
\]
in the neck.
This tilts the Gibbs free-energy landscape and reduces the entry barrier.
A convenient barrier-reduction scale is
\[
    \Delta U
    \sim
    \Phi_0 K x^\ast
    \simeq
    \frac{\Phi_0 I_{\mathrm{pk}}}{w}x^\ast,
\]
where \(x^\ast\) is a characteristic slip distance of order a few coherence lengths.
Taking \(x^\ast \sim \xi \approx 40~\mathrm{nm}\) for Nb and \(w=1~\si{\micro\meter}\), we get
\[
    \Delta U
    \approx
    \left(\frac{\Phi_0 \xi}{w}\right) I_{\mathrm{pk}}
    \approx
    8.3\times 10^{-17}\, I_{\mathrm{pk}}[\mathrm{A}]~\mathrm{J}.
\]
This yields
\[
    \Delta U \approx (3.1\text{--}5.0)\times 10^{-21}~\mathrm{J}
\]
for \(-90~\mathrm{dBm}\), and
\[
    \Delta U \approx (3.1\text{--}5.0)\times 10^{-20}~\mathrm{J}
\]
for \(-70~\mathrm{dBm}\).

Because the drive acts over many cycles, this ac Lorentz force can increase the probability for an incipient edge vortex to cross a local barrier.
The estimate above should not be interpreted as a quantitative prediction of the absolute entry field.
Rather, it shows that the resonant microwave current provides a field scale comparable to the changes in applied field over which additional neck-entry events are observed.
Thus, microwave drive can assist vortex entry at the constriction, while the actual entry field remains controlled by the local barrier, lithographic edge quality, and the nonlocal washer-field distribution.

\section{Experimental Methods and Conditions}
\subsection{Details of Magnetic-Field Measurements Using NV Centers}

\begin{figure*}[tbp]
\noindent\centering
\includegraphics[width=1.\textwidth]{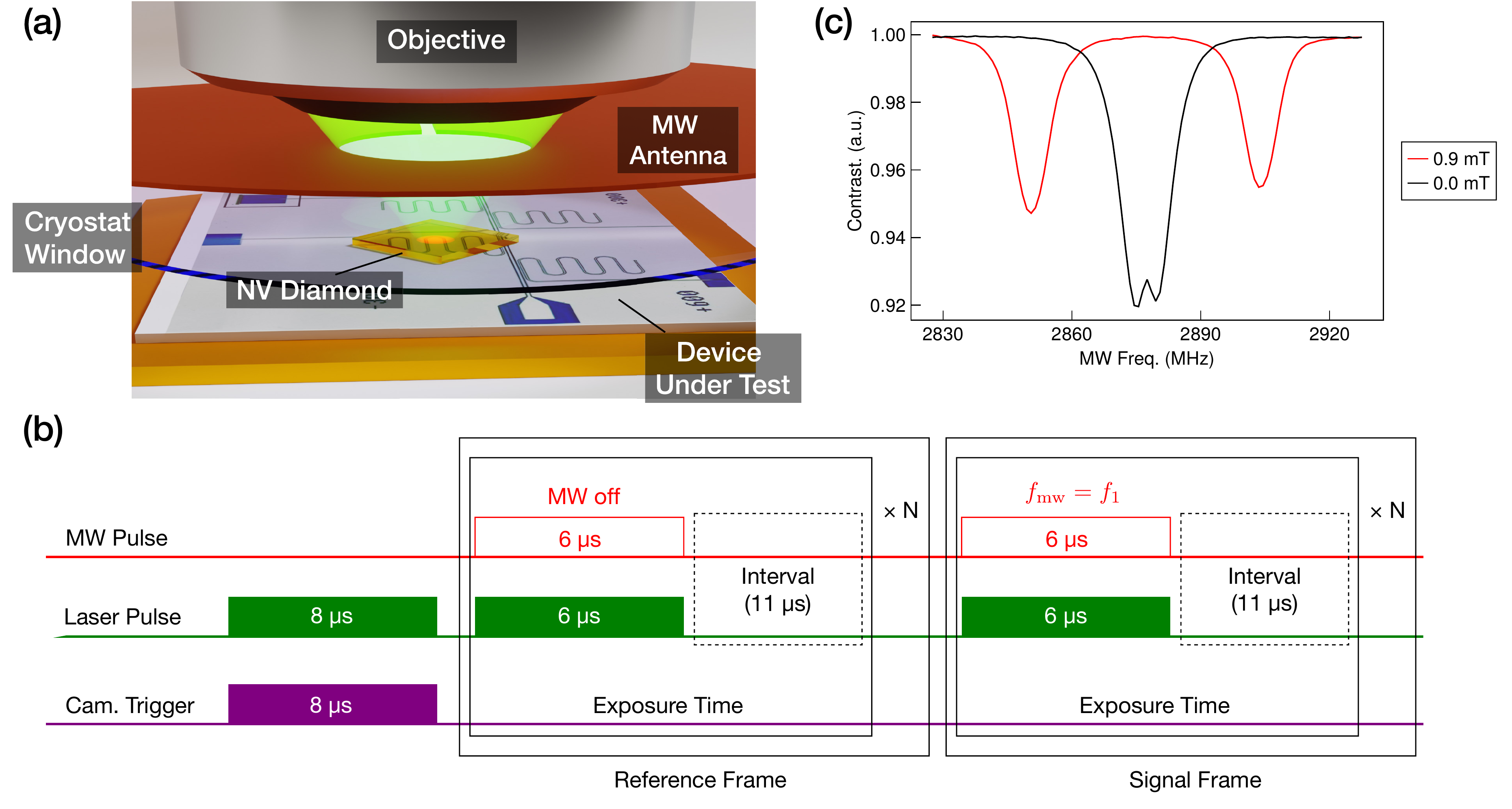}
\caption{
  (a) Schematic of our microscope.
  (b) Typical ODMR spectrum.
  (c) Diagram of the ODMR pulse protocol. Because the image sensor operates far more slowly than the time scales of NV initialization and control, we repeat the same sequence during the exposure time and read out the accumulated signal. We then sweep the microwave frequency and repeat the protocol at each point to obtain the ODMR spectrum.
}
\label{fig:qdm_diagram}
\end{figure*}

In this study we use a Quantum Diamond Microscope (QDM) based on an ensemble of nitrogen-vacancy (NV) centers [\cref{fig:qdm_diagram} (a)]~\cite{Taylor2008,Levine2019}.
As the sensor layer, we employ a perfectly oriented NV layer~\cite{SN2023,Tsuji2022} grown to a thickness of approximately \SI{1}{\micro\meter} on the diamond surface, enabling imaging of the magnetic flux density perpendicular to the superconducting thin film.
The magnetic flux density is obtained via optically detected magnetic resonance (ODMR) of the NV centers~\cite{Rondin2014}. We observe the photoluminescence (PL) intensity normalized to the presence or absence of microwave irradiation.
\Cref{fig:qdm_diagram} shows a typical ODMR spectra taken under different magnetic fields. These ODMR spectra exhibit two dips corresponding to the electronic spin resonances of the NV centers. The splitting $\Delta f$ between the resonance frequencies reflects the Zeeman effect, which is given by
\begin{equation}
\Delta f = 2\sqrt{ (\gamma_{e}B_z)^2 + E^2},
\label{eq:nv_odmr}
\end{equation}
where $B_z$ is the magnetic flux density along the NV axis, $\gamma_e=\SI{28}{\mega\hertz\per\milli\tesla}$ is the electron-spin gyromagnetic ratio, and $E$ is a strain parameter that varies with position in the crystal. At each pixel we fit the ODMR spectrum to the sum of two Lorentzian functions to extract $\Delta f$.

We also acquire a zero-field reference measurement and identify $\Delta f(B=0)$ with $E$. Using this zero-field reference and \cref{eq:nv_odmr}, we convert $\Delta f$ to $B_z$. The use of perfectly oriented NV centers simplifies the analysis: a general multi-orientation ensemble can exhibit up to eight resonance lines, which complicates fitting.
The absence of NV orientations aligned to other symmetry axes avoids contrast loss, improving sensitivity. The magnetic-field map is obtained by analyzing all \num{3200}\,$\times$\,\num{3200} CMOS pixels. 

\subsection{Cryogenic Microscope Configuration}

The sample temperature is controlled by a heater on the cold stage and monitored with a thermometer attached to the stage. [See \cref{fig:qdm_diagram} (a)]. In addition, a solenoid coil applies a static magnetic field perpendicular to the sample surface. For PL imaging of the NV centers, we excite the diamond with a green laser (\SI{515}{\nano\meter}, \SI{140}{\milli\watt}) and collect the NV PL using an image sensor (Teledyne Photometrics Kinetix) placed outside the cryostat optical window. The microwave antenna required for NV manipulation (a loop-gap-type resonator~\cite{Sasaki2016}) and the solenoid are mounted outside the optical cryostat to avoid heating.

\subsection{Pulse Protocol}
We perform the NV magnetic-field measurements using the pulse protocol shown in \cref{fig:qdm_diagram} (b). The time-averaged laser input power is therefore approximately $\SI{200}{mW} \times \frac{6}{6 + 11} \approx \SI{70.6}{mW}$. Waveforms are generated with an arbitrary waveform generator (Spectrum M4.6631-x8). The laser is the dominant heat source, which locally heats the sample and causes the actual sample temperature to deviate from the thermometer reading at the temperature-controlled stage. We keep the laser pulse waveform running continuously while acquiring data. 
We calibrate this temperature deviation by repeating the vortex measurement with different laser pulse duty cycle. We then assuming the deviation is linear to the input power, the maximum average power observing vortices (120~mW) divided by the average power for the pulse cycle we actually used (70.6~mW), which provides a rough estimate of the upper bound of the true temperature near 6.4~K.
Because the pulse-cycle period is very short (\SI{17}{\micro\second}), thermal inhomogeneities cannot follow the cycle, so the temperature rises quasi-steadily in a time-averaged sense.

\begin{figure*}[tbp]
\centering
\includegraphics[width=1.\textwidth]{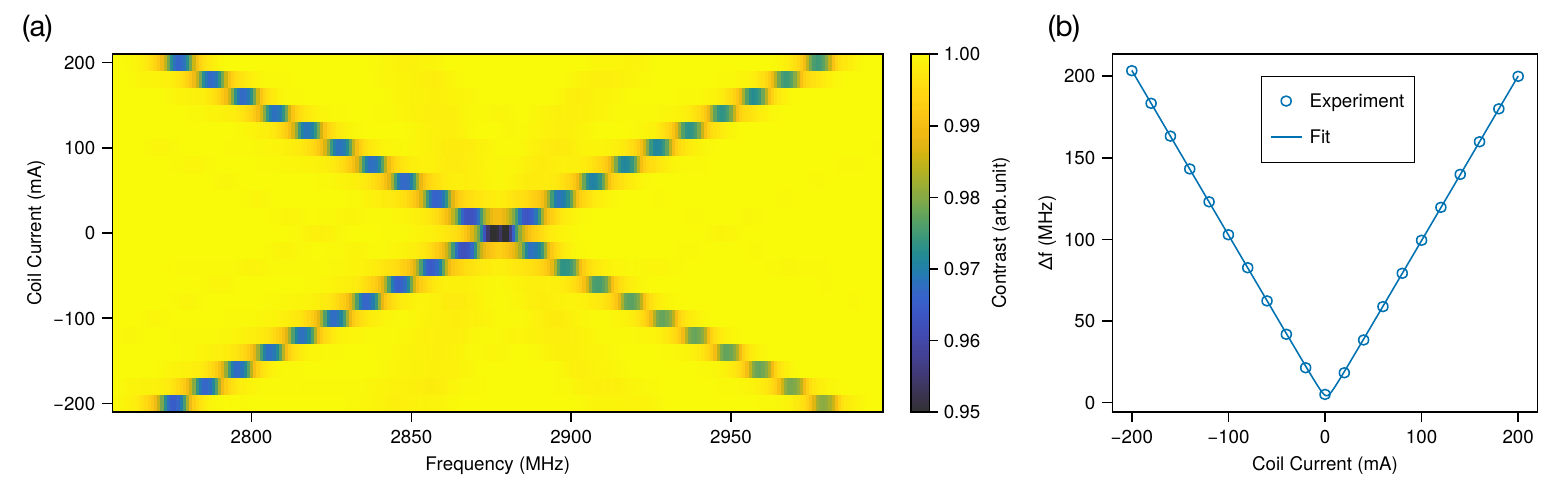}
\caption{
    Calibration of the external magnetic field using NV ODMR.
    (a) Two-dimensional intensity map of the ODMR spectra as a function of the applied coil current.
    (b) Dependence of the resonance-frequency difference $\Delta f$ on the solenoid current, extracted from (a) by fitting.
    Each value of $\Delta f$ is averaged over the whole field of view.
    We use this result to calibrate the uniform external magnetic-field bias.
}
\label{fig:coil_calib_nv}
\end{figure*}

\subsection{Magnetic-Field Calibration}

We install a solenoid coil [outside the cryostat, omitted in \cref{fig:qdm_diagram}(a) for brevity] to apply finely controlled magnetic fields to the superconductor, enabling modulation with a precision of $\lesssim \SI{1}{\micro\tesla}$.
This uncertainty arises from fluctuations of the ambient field, including the geomagnetic field, rather than from the reproducibility of the field generated by the coil.
By adjusting the current applied to the coil, we set the magnetic field, and we add a finite offset current to compensate the geomagnetic field.
Thus, the coil current $I_\text{coil}$ and the resulting $z$-directed magnetic field $B_{\mathrm{ext},z}$ satisfy
\begin{equation}
B_{\mathrm{ext},z} = \alpha \bigl(I_\text{coil} - I_0\bigr),
\label{eq:calib1}
\end{equation}
where the zero-field offset current $I_0$ (in \si{\milli\ampere}) and the current-to-field conversion factor $\alpha$ (in \si{\micro\tesla\per\milli\ampere}) are determined by applying relatively strong fields and using the NV relation below.

For calibration, we apply relatively strong magnetic fields by sweeping the current up to $200~\si{\milli\ampere}$ (corresponding to $\sim \SI{3}{\milli\tesla}$) through the same coil and measure the magnetic field at each current.
We then fit the data [\cref{fig:coil_calib_nv}(b)] with \cref{eq:calib1,eq:nv_odmr}, yielding
\begin{equation}
\alpha = 18.01(2)~\si{\micro\tesla\per\milli\ampere}
\label{eq:calib_alpha}
\end{equation}

\section*{Theoretical Estimates}
\label{sec:theory}

To perform a rough estimate of the frequency shift, we take $G \sim  \omega_0 L_{\mathrm{tot}}$, where  \(\omega_0=2\pi f_0\) and \(L_{\mathrm{tot}}\) is the total inductance of the resonator mode, which is about 3 nH, inferred from the resonator length. Taking $W\sim 1 \mu$m, $t\sim 200$ nm, we get $ \frac{\delta f}{f_0} 
\sim
\frac{{\rm Im}[Z_v(\omega_0)]}{2 G}
\sim g_I\frac{(1-\epsilon)(\omega/\omega_d)}{1+(\omega/\omega_d)^2}  \frac{\Phi_0^2} {\eta t W^2 2\pi f_0 L_{\mathrm{tot}}} \approx 3.3\cdot 10^{-6} g_I\frac{(1-\epsilon)(\omega/\omega_d)}{1+(\omega/\omega_d)^2}$.
One expects $\epsilon\ll 1$ (motivated by the low temperatures suppressing creep).  
Compared to the experimental shifts $\delta f/f\sim 0.1$ MHz/$4.8$ GHz$\sim 2\cdot 10^{-5}$, the estimate suggests a large current enhancement $g_I\sim 10$, since $\frac{(\omega/\omega_d)}{1+(\omega/\omega_d)^2}\leq 1/2$ (and likely smaller, since the pinning frequencies for Nb films thicker than 100 nm have been reported to below 1 GHz \cite{nbpinning}). 

Let us now discuss the potential magnitude of the current density enhancement in the neck region.
The current distribution in superconducting film is not uniform. According to \cite{Norris1970, Brandt1993, Zeldov1994Magnetization}, in order to cancel the self-Amp\'ere field, the current must satisfy
$$\displaystyle J=\frac{1}{\sqrt{(W/2)^2 - x^2 + \lambda_L^2}} \frac{I}{2\arcsin((W/2)/\sqrt{(W/2)^2 + \lambda_L^2})},$$ 
which is normalized to satisfy $I=\int_{-W/2}^{W/2} J dx$.
Here we allow the finite field penetration (not as a vortex form) at the edge of the order of $\lambda_L$. For vortices sitting at $100~\si{nm}$ off from the edge, $x\sim 0.4~\si{\micro m}$ which gives $J / \langle J \rangle = JW/I \approx1.1$.\\
For vortices sitting at the center, $x\sim 0.0~\si{\micro m}$ which gives $J / \langle J \rangle \approx0.68$.
For vortices sitting at the very edge, which is rather unlikely, this gives $J / \langle J \rangle = W/\lambda_L / 2\arcsin((W/2)/\sqrt{(W/2)^2 + \lambda_L^2})\sim 6.80$. The estimates above assume an infinitely long section, whereas in our experiment the ``neck" is only a small section of the resonator. Moreover, the resonator strip width is continuously reduced towards the beginning of the ``neck", such that the current distribution may be more complex. Therefore, while the estimate provided above appears within an order of magnitude of the observed frequency shift, a more rigorous modeling of the current crowding effects appears necessary to describe the frequency jumps quantitatively.

\section*{Alternative derivation of single-vortex impedance}
Here we present an alternative derivation of Eq. \eqref{eq:Zv_local} starting from equations of motion for a single vortex.
We can model the motion of vortex just as a classical harmonic oscillator, whose equation of motion is
\begin{equation}
    \eta_\mathrm{film} \dot{\bm{u}} + k_p \bm{u} = F_L = \Phi_0 J(\bm r)
\end{equation}
where $\bm r$ is the vortex location, $\bm u$ is the displacement field, $\eta_\mathrm{film}$ is the viscous drag act on a particle ($=\eta t$ in the normal coarse-grained treatment), $k_p$ corresponds to the linear coupling to the quenched random disorder (in the unit of spring constant), and $J$ is a sheet current in the unit of \si{A/m}, related to the total current as $J t=I$, $t$ being the film thickness.
Here the particle experiences a force which is proportional to J in a velocity again proportional to J.
In the frequency domain the velocity is calculated as
$$
 v(\omega) = \mu(\omega) F_L = \mu \Phi_0 J(\bm r),\quad \mu(\omega) = \frac1{\eta_\mathrm{film}}\frac{i\omega}{\omega_p + i\omega}
$$
In \cite{Coffey1991} treatment which also handles DC steady state, by adding a constant thermal force $F_\mathrm{thermal}$, $\mu$ is extended to $\displaystyle \frac1{\eta_\mathrm{film}}\frac{\epsilon\omega_p + i\omega}{\omega_p + i\omega}$.

To calculate the impedance, we use the definition of voltage from the Josephson's relation:
$$
Z_\mathrm{sv} \equiv \frac{V}{I}, \quad 
2 e V = \frac{\hbar}{ W} 
\int_0^W d y \frac{\mathrm d \phi(y,t)}{dt},
$$
where the $\frac{1}{W} 
\int_0^W d y ...$ represents averaging over the width. For a moving vortex, we have $\phi(y,t) = \phi_v(y-v t)$, where $\phi_v(0)-\phi_v(\infty) = 2\pi$. Assuming that $W$ is much larger then the vortex size, the integral over $y$ can be evaluated in this case, resulting in 
\[
V = \frac{\pi \hbar v}{e W } 
\]

Then using the expression for the velocity above, we get
\begin{align}
Z_\mathrm{sv} &= \frac{\Phi_0}{W}\mu(\omega)\Phi_0 J /I = \frac{\Phi_0^2}{W^2 t} \mu(\omega).
\end{align}

\section*{Statistics of frequency jumps and field dependence}
Resonance-frequency steps observed upon vortex entry tend to become deeper as the magnetic field increases.
This trend is visible in the representative sweep shown in Fig.~\ref{fig2}.
To examine it more closely, we extract the step heights from the $N=14$ vortex-entry events that can be unambiguously resolved in this sweep.

\begin{figure*}[tbp]
\centering
\includegraphics[width=0.5\textwidth]{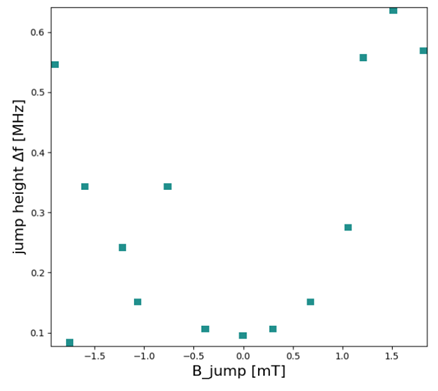}
\caption{
Scatter plot of the step height $\Delta f$ extracted for each resolved jump as a function of the magnetic field at which it occurs, $B_{\mathrm{jump}}$ (same data as used for Fig.~\ref{fig2}).
The jump amplitudes span $\Delta f \simeq 0.05$--$0.68~\mathrm{MHz}$ and tend to be larger at larger $|B_{\mathrm{jump}}|$, consistent with the qualitative observation across multiple devices/cooldowns that vortex-entry steps deepen with increasing $|B|$.}
\label{fig:statistic}
\end{figure*}

Figure~\ref{fig:statistic} shows the magnitude of each frequency jump, $|\Delta f|$, as a function of the magnetic field at which the jump occurs, $B_{\mathrm{jump}}$.
Although the step heights exhibit substantial event-to-event scatter, jumps occurring at larger $|B_{\mathrm{jump}}|$ are typically deeper than those at smaller fields.

A natural interpretation of this field dependence follows from the spatial evolution of vortex entry into the constriction.
As directly visualized by NV-center magnetometry in~\cref{fig4}, vortices do not enter the neck at random positions.
Instead, the first vortices tend to occupy regions slightly away from the narrowest part of the constriction, while vortices entering at higher fields are pushed progressively closer toward the narrowest section near the neck.
Because the microwave current density is strongly concentrated in this region, a vortex located closer to the constriction produces a larger perturbation of the resonator impedance.
As a result, vortex-entry events occurring at larger magnetic fields, which place vortices closer to the narrowest part of the neck, generate larger frequency shifts.

At the same time, the magnitude of an individual jump remains sensitive to microscopic details, including the precise vortex position relative to the current streamlines and the local pinning landscape.
These factors naturally lead to strong event-to-event variations superimposed on the overall field dependence.
The observed increase of typical jump amplitudes with $|B|$ therefore reflects the geometry-controlled nature of vortex entry in the constriction rather than a simple one-to-one correspondence between magnetic field and jump height.

\end{document}